\documentclass[english,prl,twocolumn]{revtex4-1}
\usepackage[T1]{fontenc}
\usepackage[latin9]{inputenc}
\usepackage{color}
\usepackage{amstext}
\usepackage{graphicx}

\makeatletter

\newcommand{\lyxdot}{.}

 \@ifundefined{textcolor}{}
 {%
   \definecolor{BLACK}{gray}{0}
   \definecolor{WHITE}{gray}{1}
   \definecolor{RED}{rgb}{1,0,0}
   \definecolor{GREEN}{rgb}{0,1,0}
   \definecolor{BLUE}{rgb}{0,0,1}
   \definecolor{CYAN}{cmyk}{1,0,0,0}
   \definecolor{MAGENTA}{cmyk}{0,1,0,0}
   \definecolor{YELLOW}{cmyk}{0,0,1,0}
 }

\makeatother

\usepackage{babel}
\begin{document}

\title{Genuine multipartite Einstein-Podolsky-Rosen steering}

\author{Q. Y. He$^{1,2}$ and M. D. Reid$^{1,*}$}

\affiliation{$^{\text{1}}$Centre for Quantum Atom Optics, Swinburne University
of Technology, Melbourne, 3122 Australia\\
$^{\text{2}}$State Key Laboratory of Mesoscopic Physics, School of
Physics, Peking University, Beijing 100871 China \inputencoding{latin1}{}\\
\inputencoding{latin9}$^{*}$\inputencoding{latin1}{mdreid@swin.edu.au}}
\begin{abstract}
We develop the concept of genuine $N-$partite Einstein-Podolsky-Rosen
(EPR) steering. This nonlocality is the natural multipartite extension
of the original EPR paradox. Useful properties emerge that are not
guaranteed for genuine multipartite entangled states. In particular,
there is a close link with the task of one-sided device-independent
quantum secret sharing. We derive inequalities to demonstrate multipartite
EPR steering for Greenberger-Horne-Zeilinger (GHZ) and Gaussian continuous
variable (CV) states in loophole-free scenarios. 
\end{abstract}
\maketitle
Bell's seminal work showed that quantum mechanics is not equivalent
to any local hidden variable theory (LHV) \cite{Bell}, but this work
was a study of nonlocality between two particles only. Svetlichny
asked whether quantum mechanics could exhibit a \emph{genuine} three-body
nonlocality \cite{svetlichny}, in which case the nonlocality cannot
be simulated by any nonlocality that might exist between only two
bodies. These ideas are crucial to understanding the full nature of
the transition from the quantum to the classical regime \cite{svet recent int,ghose,collins localnonlocal-2,svetexp,bancalnew,acinoperframwork,seevinck}.
Seevinck, Collins and co-workers \cite{collins localnonlocal-2,seevinck}
revealed that $N$-party Greenberger-Horne-Zeilinger (GHZ) states
can exhibit genuine Bell nonlocality among $N$ sites, and experiments
have reported violation of Svetlichny inequalities using GHZ states
\cite{svetexp,gHZ}. The experimental violation however was limited
to $N=3$, and to systems of only one qubit (photon) per site.

Our knowledge of multipartite entanglement on the other hand is much
more established. Experimental signatures have been developed, for
both continuous variable (CV) \cite{cvsig} and qubit systems \cite{w state,papp,bancalgendient,threeent}.
There has been experimental evidence in both cases \cite{papp,wine,aokicv},
with the generation of fourteen entangled qubits in ion-traps \cite{14blatt}
and recent reports of CV entanglement of up to eight light modes \cite{bachlam}.
However, entanglement does not demonstrate nonlocality \cite{werner-1,hw-1,jones},
and it is widely appreciated that the detection of Bell nonlocality
is far more challenging \cite{bell loop}. Whether the observation
of genuine $N$-partite Bell nonlocality is possible for systems of
very high dimension or for CV measurements is not yet fully understood
\cite{svet recent int}. Despite this, there is an increasing awareness
that nonlocality is not only fundamentally significant, but can be
specifically required for certain quantum information tasks \cite{hw-1,cryp,cv cry,eprappligrangcopy,jones,bellsecu,bancalgendient}. 

In this Letter, we investigate an intermediate type of genuine $N$-body
nonlocality. As it is potentially less susceptible to noise and decoherence
than Bell nonlocality, it is therefore more accessible to experiment.
We consider genuine multipartite forms of Einstein-Podolsky-Rosen
(EPR) steering. Steering has only recently been identified as a distinct
type of nonlocality \cite{hw-1,saun,jones}, different to both entanglement
and Bell's nonlocality, and is realised in experiments that reveal
an EPR paradox \cite{einstein,rmp}. Work by Wiseman and co-workers
\cite{hw-1,jones} formalised Schrodinger's concept of ``steering'',
that an observer can apparently instantaneously influence a distant
system, by making local measurements. Multipartite EPR steering has
been studied for qubits \cite{cavalmulti-1} and qudits \cite{mulitqudits}.
However, this work did not examine genuine multipartite nonlocality,
in which the nonlocality is necessarily shared among all observers. 

We show that it is possible to obtain genuine multipartite EPR steering
in very different sorts of systems to those so far predicted for multipartite
Bell nonlocality. To date, EPR steering has been verified at very
high detection efficiencies for CV Gaussian optical systems \cite{rmp,cvsteerrecent,onewaysteer}
and without detection loopholes for photons \cite{smithsteerxp,bwzeil,loopholesteer},
but the focus has been on the bipartite case.

Here, we formalise the meaning of genuine multipartite EPR steering,
and derive criteria to detect it. We show how to verify $N$-partite
steering for GHZ states, both in discrete and CV Gaussian systems,
giving efficiency bounds to do so conclusively. Our work therefore
opens up possibilities to demonstrate an $N$-partite EPR nonlocality
($N>2$) unambiguously for qubit sites, whether by using photons \cite{svetexp,smithsteerxp,loopholesteer}
or ions \cite{wine,14blatt}, and to test the existence of the strongest
form of nonlocality so far predicted to distribute over many sites
with systems in the continuous (CV) limit. Further, we prove results
for multipartite EPR steering, that are useful to multi-party quantum
communication protocols \cite{quantum internet,secretsh,secretsh-1},
and are not guaranteed by multipartite entanglement. 

\emph{Genuine $N$-partite nonlocality:} We consider $N$ spatially
separated systems at sites $j=1,...,N$, and ask how to derive criteria
for genuine\emph{ }$N$-party nonlocality, so that we can conclude
nonlocality to be shared among \emph{all} $N$ parties. The strongest
form of nonlocality is Bell's nonlocality\emph{, }in which all Local
Hidden Variable (LHV) models are falsified \cite{Bell}. Denoting
the hidden variables that specify the predetermined nature of the
system by the set $\{\lambda\}$, all LHV theories will imply the
fully separable LHV model $\langle\prod_{j=1}^{N}X_{j}\rangle=\int_{\lambda}d\lambda P(\lambda)\prod_{j=1}^{N}\langle X_{j}\rangle_{\theta_{j},\lambda}$.
The $j-1$ factorisations in the integrand are justified based on
the assumptions of locality and no-signalling between all sites $j$.
Here $X_{j}$ are the possible results for a measurement $\hat{X}_{j}$
at site $j$, $\langle X_{j}\rangle_{\theta_{j},\lambda}$ is the
expected value of $X_{j}$ for a given set $\{\lambda\}$ where $\theta_{j}$
denotes the choice of measurement at site $j$, and $P(\lambda)$
is the hidden variable probability distribution function. Bell's nonlocality
is demonstrated when the LHV model fails. 

Genuine $N$-party Bell nonlocality\emph{ }can be tested using a method
pioneered by Svetlichny \cite{svetlichny,collins localnonlocal-2,seevinck}.
We construct a hybrid local-nonlocal model in which Bell nonlocality\emph{
}exists, but only if shared among $N-1$ or fewer parties.\emph{ }Thus,
the fully separable LHV model becomes only partially separable, with
separability retained between any two groups $A$ and $B$ of $ $$N-k$
and $k$ ($k\leq N/2$) parties respectively. We label the possible
ways of splitting the sites into two such groups by the index $s$.
 The Svetlichny model is 
\begin{eqnarray}
\langle\prod_{j=1}^{N}X_{j}\rangle & = & \sum_{s}P_{s}\int_{\lambda}d\lambda P_{s}(\lambda)\nonumber \\
 &  & \langle\prod_{j\in A_{s}}X_{j}\rangle_{\{\theta_{j}\},\lambda}\times\langle\prod_{j\in B_{s}}X_{j}\rangle_{\{\theta_{j}\},\lambda}\label{eqn:sepave-1-1}
\end{eqnarray}
where $\sum_{s}P_{s}=1$. Failure of (1) implies genuine $N$-partite
Bell nonlocality. Recently, Gallego et al \cite{acinoperframwork}
and Bancal et al \cite{bancalnew} have revealed that the Svetlichny
definition, which assumes fully ``bi-local'' (BL) correlations,
is strictly stronger than necessary to confirm genuine $N$-partite
nonlocality. Surprisingly, the bipartition $\{A_{s},B_{s}\}$ can
admit a Bell nonlocality, if all types of signalling between the parties
of a group are allowed. By relaxing the assumptions of the model,
to allow only ``time-ordered bi-local correlations'' (TOBL), a more
sensitive test can be obtained. 

We next consider the three different types of nonlocality $-$ Bell
nonlocality, steering, and entanglement $-$ that may exist between
two sites, as introduced by Wiseman and co-workers \cite{hw-1,cavalsteer,jones}.\emph{
}Following those authors, the fully separable LHV model becomes a
quantum separable model when there exists a local quantum density
operator $\rho_{j}^{\lambda}$ such that $\langle X_{j}\rangle_{\lambda}=Tr(\rho_{j}^{\lambda}X_{j})$,
for each $j$. In this case, the system is described by a fully separable
density matrix $\rho$ and failure of the model implies entanglement\emph{.
}To test for genuine $N$-partite entanglement\emph{, }a partially
separable model is used \cite{cvsig,w state}, where $\rho=\sum_{s}P_{s}\int_{\lambda}d\lambda P_{s}(\lambda)\rho_{A_{s}}^{\lambda}\rho_{B_{s}}^{\lambda}d\lambda$
and $\rho_{A_{s}}^{\lambda}$ is a density operator, not necessarily
factorisable, for the group $A_{s}$ (similarly $\rho_{B_{s}}^{\lambda}$).
This is equivalent to the hybrid local-nonlocal model (\ref{eqn:sepave-1-1})
but with the further constraint that moments for $A_{s}$ and $B_{s}$
each arise from a quantum density matrix, $\rho_{A_{s}}^{\lambda}$and
$\rho_{B_{s}}^{\lambda}$, respectively. The failure of all such models
demonstrates genuine $N$-partite entanglement \cite{bancalgendient}.

\begin{figure}
\begin{centering}
\textcolor{black}{}
\par\end{centering}

\begin{centering}
\includegraphics[width=0.6\columnwidth]{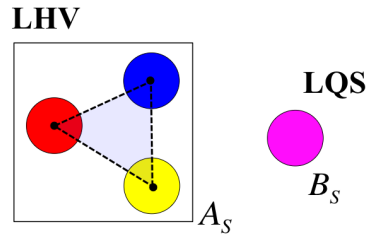}
\par\end{centering}

\caption{Depiction of one of the bipartitions $\{A_{s},B_{s}\}_{st}$ of the
Hybrid Local Hidden State (LHS) local-nonlocal model for $N=4$. Here
three sites can share a Bell nonlocality, but four cannot. The group
$A_{s}$ ``steers'' system $B_{s}$, if this model fails, when it
is also constrained that $B_{s}$ be consistent with a Local Quantum
State.}
\end{figure}

Now we turn to the case of steering. Following Ref. \cite{hw-1},
we impose the asymmetric constraint on the model (\ref{eqn:sepave-1-1})
that there exists a quantum density operator $\rho_{B_{s}}^{\lambda}$
for the group of sites labelled $B_{s}$, but not\emph{ }for the group
labelled $A_{s}$. Failure of this model, called a Local Hidden State
(LHS) model, demonstrates\emph{ }``steering'' of system $B_{s}$
by measurements performed on $A_{s}$ (Fig. 1) \cite{hw-1} (we will
abbreviate, to say ``$A_{s}$ steers $B_{s}$''). Such steering
can be confirmed through the violation of EPR-steering inequalities
that are closely associated with the EPR paradox \cite{cavalsteer,saun}.
A hierarchy of nonlocality is implied by the definitions. The local
quantum state (LQS) description $\rho_{A_{s}}^{\lambda}$ is a particular
example of a local hidden variable (LHV) one. Hence, Bell nonlocality
between two groups $A_{s}$ and $B_{s}$ implies both steering and
entanglement, and steering implies entanglement (but not Bell nonlocality).
Unlike the other two nonlocalities, EPR steering is directional\emph{:}
that $A$ can steer $B$ does not imply that $B$ can steer $A$ \cite{onewaysteer}.

A definition of genuine multi-partite steering follows naturally.\emph{
}Genuine tripartite steering exists iff it can be shown that a steering
nonlocality is necessarily shared among all $3$ sites. This means
that the system cannot be described by any state in which steering
is shared among $ $two sites only. In this paper, we say two parties
``share steering'' if (at least) one can be shown to steer the other.
The definition must be consistent with the operational definitions
of Refs. \cite{bancalnew,acinoperframwork}, namely, that steering
shared among $N$ parties cannot be created by local operations and
allowed classical communication (LOCC), even when $N-1$ parties collaborate.
It will become necessary to constrain the collective hidden variable
groups $A_{s}$ so that they satisfy no-signalling principles \cite{bancalnew,acinoperframwork}.

However, we are able to use the partitioning of the model (\ref{eqn:sepave-1-1})
to arrive at conditions that suffice to detect genuine multipartite
steering. We now prove four results, which lead to simple criteria,
and give properties that make multipartite EPR steering useful.

\textbf{Result (1}):\emph{ Hybrid LHS local-nonlocal model: }The failure
of the model (\ref{eqn:sepave-1-1}) where each group $B_{s}$ consists
of one LQS site and group $A_{s}$ consists of $2$ sites is sufficient
to demonstrate genuine tripartite steering. 

\emph{Proof}: It has been proved that the set of TOBL states are strictly
contained within the set of BL states \cite{bancalnew,acinoperframwork}.
Hence, falsification of all Svetlichny models (\ref{eqn:sepave-1-1})
is sufficient to demonstrate genuine tripartite Bell nonlocality.
The possible bipartitions $\{A_{s},B_{s}\}$ are $\{\{1,2\},3\}$,
$\{\{1,3\},2\}$ and $\{\{2,3\},1\}$. The model allows for Bell nonlocality
between the sites of $A_{s}$. Hence, by the hierarchy of nonlocality,
all ways in which steering can be shared between any two sites are
described by the model. The failure of the model where site $B_{s}$
is a LQS is sufficient to imply steering, though not Bell nonlocality,
between the groups $A_{s}$ and $B_{s}$. Thus, genuine tripartite
steering is demonstrated if this model is falsified. The extension
for larger $N$ is discussed in the Supplemental Material \cite{proofepr}.

To falsify the model of Result (1), we need to rule out all \emph{mixtures
}of the relevant bipartitions (which we denote $\{A_{s},B_{s}\}_{st}$)
that account for the way steering can be shared between $2$ parties.
The next result tells us how to rule out that the system can be described
by any \emph{one }of these bipartitions. This will prove genuine tripartite
steering (and more generally $N$-partite steering), if we are constrained
to pure state models, where mixtures of different bipartitions are
not possible (a property that holds for pure quantum states). First,
we introduce a definition: we demonstrate ``\emph{collective $N$-partite
steering} of a system $B$'', if it is shown that the steering of
$B$ by a group of $N-1$ parties $A$ cannot also be demonstrated
by the measurements of fewer than $N$ parties. As might be expected,
we find a close relationship between collective $N$-partite steering,
which always involves $N$ parties, and genuine $N$-partite steering.

\textbf{Result (2):} \emph{Pure-state genuine $N$-partite steering}:
(a) All bipartitions $\{A_{s},B_{s}\}_{st}$ of the model are negated,
if it is shown that each group $\{A_{s}\}$ collectively steers $B_{s}$.
(b) For the tripartite case, it is sufficient to show that each party
can be steered by one or both of the other two. 

\emph{Proof:} (a) Take by example $N=4$. The possible bipartitions
$\{A_{s},B_{s}\}$ are $\{\{1,2,3\},4\},$$\{\{1,2\},\{3,4\}\}$,
and all permutations. Steering of any one party $B_{s}$ by $A_{s}$
negates the first type of bipartition. The second type of partition
is negated by the steering of $B_{s}$, that cannot be described as
steering by only one other party. (b) Take each of $\{\{1,2\},3\}$,
$\{\{1,3\},2\}$ and $\{\{2,3\},1\}$. The first is negated if 3 is
steered, the second if $2$ is steered, the third if $1$ is steered.

The Result (2) shows that collective steering is a type of genuine
multipartite steering. If we demonstrate multipartite collective steering,
then we confirm genuine multipartite steering, for the case of pure
states. 

The usefulness of this sort of genuine multipartite steering can be
understood, once we realise the connection between steering and the
EPR paradox \cite{hw-1,cavalsteer,einstein}. It has been shown in
Refs. \cite{cavalsteer,cavalmulti-1,saun} that steering of a single
spin-$1/2$ qubit $B$ is confirmed, if a group of parties $A$ can
infer, with sufficient accuracy, \emph{both} of the Pauli spin components,
$\sigma_{x}^{B}$ and $\sigma_{y}^{B}$. (They use different measurements
for each inference). Similarly, steering of a single harmonic oscillator
is confirmed, if the parties $A$ can infer accurately the values
of \emph{both} position and momentum, $x_{B}$ and $p_{B}$. Specifically,
for any quantum state at $B$, the quantum uncertainty relations $(\Delta\sigma_{x}^{B})^{2}+(\Delta\sigma_{y}^{B})^{2}\geq1$
and $\Delta x_{B}\Delta p_{B}\geq1$ must hold \cite{hoftoth} (we
assume appropriate scaling). We confirm the steering of $B$ by $A$,
if 
\begin{equation}
S_{B|A}\equiv\Delta_{inf,A}x_{B}\Delta_{inf,A}p_{B}<1\label{eq:eprcv}
\end{equation}
 where $\Delta_{inf,A}x_{B}$ is the uncertainty in the prediction
$x_{pred}$ of $x_{B}$ based on local measurements on system $A$
\cite{hw-1,cavalsteer,eprr-2}. Alternatively, we confirm steering
if 
\begin{equation}
\mathcal{S}_{B|A}\equiv(\Delta_{inf,A}\sigma_{x}^{B})^{2}+(\Delta_{inf,A}\sigma_{y}^{B})^{2}<1\label{eq:eprspin}
\end{equation}
These inequalities have been used to confirm the EPR paradox, for
bipartite systems \cite{rmp}.

Now we can establish a link with the quantum information protocol
of ``quantum secret sharing'' \cite{secretsh,secretsh-1}. Suppose
the parties at $A$ are shown to steer $B$, by measurements that
reveal a reduced noise on the inferences, so that $S_{B|A}<1$. If
it is known that group $A$ collectively steers $B$, then (by definition)
the uncertainty product $S_{B|A}$ based on \emph{any} measurements
made by \emph{fewer} parties at $A$ must exceed $1$. Thus, collective
$N$-party steering provides the resource for ``$N$-party quantum
secret sharing'', whereby $N-1$ parties must collaborate in order
to deduce, by measurements on their systems, the value of the amplitude/
qubit of system $B$. 

This motivates us to prove another useful result. A monogamy of EPR
steering holds, that guarantees minimum noise levels on any inferences
that could be made by ``eavesdropping parties''. The noise levels
are a direct consequence of the quantum uncertainty relation of system
$B$ \emph{only}, and therefore provide ``one-sided device-independent''
security, whereby no assumptions are made about the exact nature of
the strategies of the eavesdropping parties \cite{eprappligrangcopy,bellsecu}.
Moreover, the amount of noise incurred is directly determinable from
the degree of steering measured by the collaborating parties, $A$
and $B$.

\textbf{Result (3): }\emph{Steering and security:} If it can be proved
by violation of an EPR steering inequality involving two observables
at each site that group $A$ steers $B$, then there can be no third
group $C$ independent of $A$ that can also violate the same inequality.\textcolor{black}{{}
}\textcolor{red}{}\textcolor{black}{In particular, $S_{B|A}S_{B|C}\geq1$.}

\emph{Proof:} Consider the EPR steering inequality $S_{B|A}<1$ defined
above. Group $A$ performs measurements to predict $x_{B}$ ($p_{B}$)
with uncertainty $\Delta_{inf,A}x_{B}$ ($\Delta_{inf,A}p_{B}$) while
simultaneously $C$ can perform measurements to predict $p_{B}$ ($x_{B})$
with uncertainty $\Delta_{inf,C}p_{B}$ ($\Delta_{inf,C}x_{B}$).
The Heisenberg uncertainty principle implies \textcolor{black}{$S_{B|A}S_{B|C}\geq1.$
More generally, the proofs follow because the state of $B$ conditioned
on the joint outcomes of A and C is a quantum state.} 

We return to the fundamental and challenging problem of detecting
\emph{genuin}e $N$-partite steering. To do this, without the pure
state assumption, we need to negate all \emph{mixtures} of the possible
bipartitions $\{A_{s},B_{s}\}_{st}$, that describe the possible ways
to share steering among fewer than $N$ parties. For some systems,
this can be done by violation of a single EPR steering inequality,
and we give derivations in the Supplemental Materials \cite{proofepr}.
Generally, however, it is enough to demonstrate ``strong'' EPR steering
of each party, by the others. 

\textbf{Result (4):} \emph{Genuine tripartite steering:} Suppose $S_{i}\equiv S_{i|\{jk\}}<1$
indicates steering of party $i$ by the other parties $\{j,k\}$,
where $S_{i|\{jk\}}$ is either the product or the sum of conditional
inference variances, as defined for (\ref{eq:eprcv}) and (\ref{eq:eprspin}).
Genuine tripartite steering is confirmed whenever $S_{1}+S_{2}+S_{3}<1$. 

\emph{Proof:} Suppose the system is a mixture of the three bipartitions
$\{\{1,2\},\{3\}\}_{st}$, $\{\{1,3\},\{2\}\}_{st}$ and $\{\{2,3\},\{1\}\}_{st}$,
which we label $s=3,2,1$ respectively, with relative probabilities
$P_{s}$. For any such mixture: $S_{i}\geq\sum_{s}P_{s}S_{i,s}$ where
$S_{i,s}$ denotes the value $S_{i}$ for a system in bipartition
$s$ \cite{hoftoth}. By definition of steering, $S_{i,i}\geq1$.
Hence, $S_{i}\geq P_{i}$, and the result follows on using $\sum_{s}P_{s}=1$.

\emph{$N$- partite GHZ qubit states:} The $N$- spin GHZ state $\frac{1}{\sqrt{2}}\{|\uparrow\rangle^{\otimes N}-|\downarrow\rangle^{\otimes N}\}$
predicts genuine $N$-partite steering. Here $|\downarrow\rangle_{j}$,
$|\uparrow\rangle_{j}$ are eigenstates of the Pauli spin $\sigma_{z}^{(j)}$
of the $j$th particle. For example, for $N$ odd, the GHZ state
is an eigenstate of $\sigma_{x}^{(N)}\prod_{j=1}^{N-1}\sigma_{y}^{(j)}$
(and all other products arising from the permutations among the $N$
sites) \cite{mermin,gHZ}. Thus, any group $A$ of $N-1$ observers
is able to predict the outcome of the spin $\sigma_{x}^{(N)}$ of
the $N$th particle, by measuring $\sigma_{x,pred}^{(N)}=(-1)^{(N+1)/2}\prod_{j=1}^{N-1}\sigma_{y}^{(j)}$.
In a similar way, the spin $\sigma_{y}$ of the $N$th particle can
be predicted if the $N-1$ observers measure the spin product $\sigma_{y,pred}^{(N)}=(-1)^{(N+1)/2}\sigma_{x}^{(N-1)}\prod_{j=1}^{(N-2)}\sigma_{y}^{j}$.
The variances $(\Delta_{inf,A}\sigma_{x/y}^{(N)})^{2}$ are zero for
the GHZ state.  In fact, from (\ref{eq:eprspin}), we see that
measurement of 
\begin{equation}
\mathcal{S}_{N}=(\Delta(\sigma_{x}^{(N)}-\sigma_{x,pred}^{(N)}))^{2}+(\Delta(\sigma_{y}^{(N)}-\sigma_{y,pred}^{(N)}))^{2}<1\label{eq:eprsteeruncertainty-1}
\end{equation}
implies EPR steering of the $N$-th spin. For $N=3,$ a measurement
of $S_{j}<1/3$ for each $j=1,2,3$ would be sufficient to confirm
genuine tripartite steering. Since two observables, $\sigma_{x}$
and $\sigma_{y}$, are involved, Result (3) applies, to ensure that
the inferences of any third group of observers $C$ cannot satisfy
(\ref{eq:eprsteeruncertainty-1}). 

EPR steering inequalities with three observables have recently enabled
verification of steering in a ``loophole-free'' way without fair
sampling assumptions \cite{loopholesteer,smithsteerxp,bwzeil}. Motivated
by this, we have derived a multipartite EPR steering inequality for
$3$ spins: $(\Delta_{inf,A}\sigma_{x}^{B})^{2}+(\Delta_{inf,A}\sigma_{y}^{B})^{2}+(\Delta_{inf,A}\sigma_{z}^{B})^{2}<2$.
We find the inequality is satisfied for GHZ states provided the overall
detection efficiency of group $A$ exceeds $1/3$, making loophole-free
photonic demonstration experimentally feasible. Details are given
in the Supplemental Materials \cite{proofepr}.

\emph{CV GHZ states}:\textbf{ }Consider $N$ harmonic oscillators
(fields) at sites $j$, with boson operators $a_{j}$. The quadrature
amplitudes $x_{j}$, $p_{j}$ are given by $a_{j}=x_{j}+ip_{j}$.
A tripartite CV GHZ state is a simultaneous eigenstate of $x_{k}-x_{j}$
and $p_{1}+p_{2}+p_{3}$ with eigenvalues $0$ \cite{aokicv}. For
such a state: \cite{olsen tripr}:
\begin{eqnarray}
(\Delta_{inf,\{km\}}x_{j})^{2} & = & (\Delta(x_{j}-x_{k}))^{2}=(\Delta(x_{j}-x_{m}))^{2}=0\nonumber \\
(\Delta_{inf,\{km\}}p_{j})^{2} & = & (\Delta(p_{j}+(p_{k}+p_{m})))^{2}=0\label{eq:cvghzeqn}
\end{eqnarray}
where $j,k,m=1,2,3$ and $j\neq k\neq m$. Measurement of
\begin{equation}
S_{j}\equiv\Delta(x_{j}-x_{k})\Delta(p_{j}+(p_{k}+p_{m}))<1\label{eq:cvsteer}
\end{equation}
will confirm EPR steering of system $j$, and, by Result (4), $S_{1}+S_{2}+S_{3}<1$
will confirm genuine tripartite steering.

\begin{figure}
\begin{centering}

\par\end{centering}

\includegraphics[width=0.6\columnwidth]{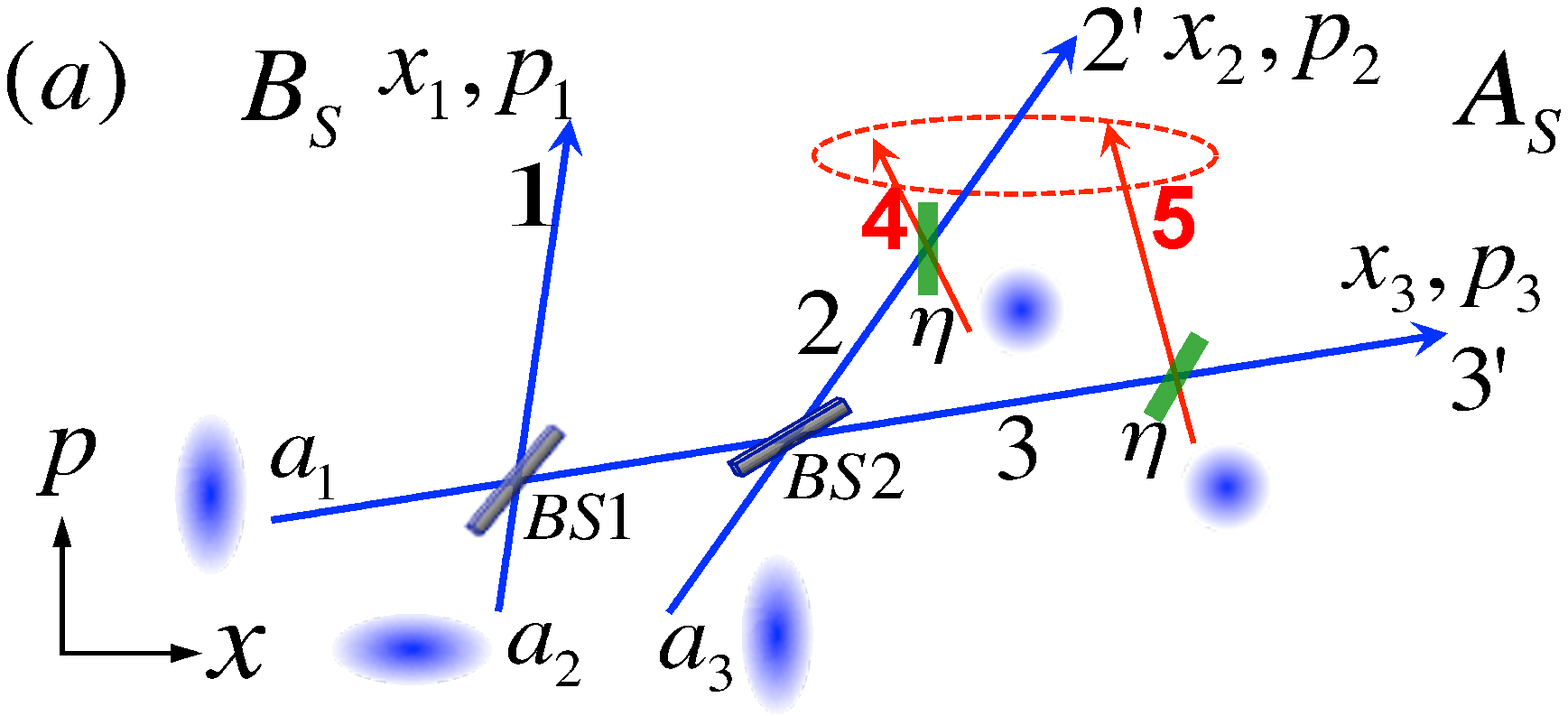}

$\ \ $$\ $\includegraphics[width=0.6\columnwidth]{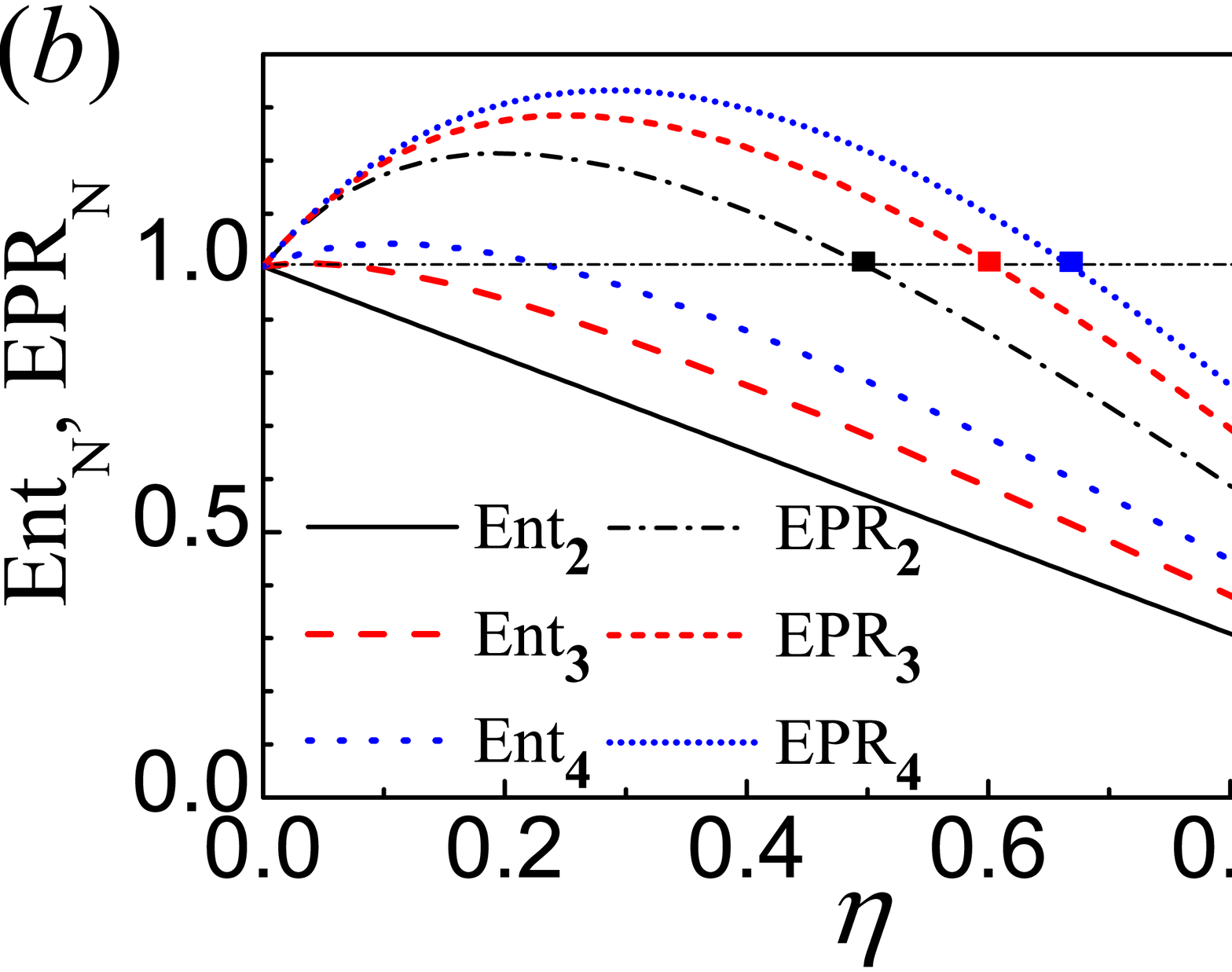}

\caption{(a) Genuine CV tripartite entanglement for beams $\{1,2,3\}$ can
be generated via squeezed states and beam splitters (BS) \cite{aokicv}.
(b) Genuine $ $tripartite entanglement exists for $\{1,2',3'\}$
when $Ent_{3}<1$ whereas EPR steering of $\{1\}$ by $\{2',3'\}$
is confirmed when $EPR_{3}=\Delta_{inf,A'}x_{1}\Delta_{inf,A'}p_{1}<1$.
Here, $\eta$ is the efficiency of transmission of the eavesdropping
beam splitters. Cases $N=2,4$ are also shown. The loss of steering
reveals the eavesdropper. \textcolor{red}{}\label{fig:EPRsteering}
 }
\end{figure}

\emph{Entanglement versus EPR steering:}$ $ \textcolor{black}{We
point out that the properties discussed in this paper do not hold
generally for multipartite entangled states. For example, consider
the CV GHZ state depicted in Fig. 2a. This state shows collective
steering, illustrated by the exact predictions for $x_{1}$ and $p_{1}$
by systems $A=\{2,3\}$, so that $\Delta_{inf,A}x_{1}\Delta_{inf,A}p_{1}\rightarrow0$.
If the beams 2 and 3 are coupled to 50:50 beam splitters, the collective
steering necessarily vanishes}. Two symmetric sets of beams exist
in this case: $A'=\{2',3'\}$ and an eavesdropping set, $E=\{4,5\}$.
Hence, by Result (3), \textcolor{black}{$\Delta_{inf,A'}x_{1}\Delta_{inf,A'}p_{1}=\Delta_{inf,E}x_{1}\Delta_{inf,E}p_{1}\geq1$}.
Yet, both sets remain genuine tripartite entangled with beam $1$
(Fig. 2b), as measurable using the two-observable inequalities of
Ref. \cite{cvsig,bow}. Hence, Result (3) does not hold for multipartite
entanglement.

\emph{Conclusion: }We have introduced the genuine tripartite EPR steering
nonlocality, established its potential importance as a resource for
secure quantum communications, and derived criteria that can be applied
to current experiments. The observation of multipartite EPR steering
in any of these systems would seem very feasible. 

We thank A. Samblowski, S. Armstrong, Ping Koy Lam and P. Drummond
for stimulating discussions. The work was suppported by Australian
ARC Discovery Project and DECRA grants. Q. H. thanks support from
China NNSF Grant No. 11121091.


\begin{thebibliography}{References}
\bibitem{Bell}J. S. Bell, Physics \textbf{1,} 195 (1964).

\bibitem{svetlichny}G. Svetlichny, Phys. Rev. D \textbf{35}, 3066
(1987).

\bibitem{svet recent int} \textcolor{blue}{}J. D. Bancal et al.,
Phys. Rev. Lett. \textbf{106}, 020405 (201\textcolor{black}{1). G.
Adesso and S. Piano, arXiv:1307.3288v2.}

\bibitem{ghose}S. Ghose et al., Phys. Rev. Lett. \textbf{102}, 250404
(2009).

\bibitem{collins localnonlocal-2} D. Collins et al., Phys. Rev. Lett.
\textbf{88}, 170405 (2002).

\bibitem{seevinck}M. Seevinck and G. Svetlichny, Phys. Rev. Lett.
\textbf{89}, 060401 (2002).

\bibitem{acinoperframwork} R. Gallego et al., Phys. Rev. Lett., \textbf{109},
070401 (2012).

\bibitem{bancalnew}J. Bancal et al., Phys Rev A \textbf{88}, 014102
(2013).

\bibitem{gHZ}D. M. Greenberger, M. A. Horne and A. Zeilinger, in
``Bell\textquoteright{}s Theorem, Quantum Theory, and Conceptions
of the Universe'' (Kluwer, Dordrecht, 1989), p. 69.

\bibitem{svetexp}J. Lavioe et al., New J. Phys \textbf{11}, 073051
(2009)\textcolor{blue}{. }\textcolor{black}{H. Lu et al., Phys. Rev.
A }\textbf{\textcolor{black}{84}}\textcolor{black}{, 012111 (2011)}\textcolor{blue}{.}

\bibitem{cvsig}P. van Loock and A. Furusawa, Phys. Rev. A \textbf{67},
052315 (2003). 

\bibitem{w state} R. Horodecki et al., Rev. Mod. Phys. \textbf{81},
865 (2009). O. Guhne and M. Seevinck, New J. Phys. \textbf{12,} 053002
(2010).

\bibitem{bancalgendient}J. Bancal et al., Phys. Rev. Lett. \textbf{106,}
250404 (2011). 

\bibitem{papp}S. B. Papp et al., Science \textbf{324}, 764 (2009). 

\bibitem{threeent}L. K. Shalm et al., Nat. Phys. \textbf{9}, 19 (2013). 

\bibitem{wine}D. Liebfried et al., Nature, \textbf{438}, 639 (2005). 

\bibitem{aokicv}T. Aoki et al., Phys. Rev. Lett. \textbf{91}, 080404
(2003). B. Hage et al., \textit{\emph{Phys. Rev. A}} \textbf{81},
062301 (2010). P. van Loock and S. L. Braunstein, Phys. Rev. Lett.
\textbf{84}, 3482 (2000). P. van Loock and S.L. Braunstein, Phys.
Rev. A \textbf{63}, 022106 (2001).

\bibitem{14blatt}T. Monz et al., Phys. Rev. Lett. \textbf{106}, 130506
(2011).

\bibitem{bachlam}S. Armstrong et al., Nature Commun. \textbf{3},
1026 (2012). 

\bibitem{werner-1}\textcolor{black}{R. F. Werner, Phys. Rev. A }\textbf{\textcolor{black}{40}}\textcolor{black}{,
4277 (1989). }

\bibitem{hw-1}H. M. Wiseman, S. J. Jones, and A. C. Doherty, Phys.
Rev. Lett. \textbf{98}, 140402 (2007).

\bibitem{jones}S. Jones et al., Phys. Rev. A \textbf{76}, 052116
(2007). 

\bibitem{bell loop}A. Cabello et al., Phys. Rev. Lett. \textbf{101},
120402 (2008).

\bibitem{cryp}A. Ekert, Phys. Rev. Lett. \textbf{67}, 661 (1991).

\bibitem{bellsecu}A. Acin, N. Gisin and L. Masanes, Phys Rev. Lett.
\textbf{97}, 120405 (2006). A. Acin et al., Phys. Rev. Lett. \textbf{98},
230501 (2007).

\bibitem{cv cry} T. C. Ralph, Phys. Rev. A \textbf{61}, 010303(R)
(1999); \textcolor{black}{V. Scarani et al., Rev. Mod. Phys. }\textbf{\textcolor{black}{81}}\textcolor{black}{,
1301 (2009);} Lars S. Madsen et al., Nature Commun. \textbf{3}, 1083
(2012).\textcolor{red}{{} }

\bibitem{eprappligrangcopy}C. Branciard et al., Phys. Rev. A \textbf{85},
010301(R) (2012).

\bibitem{saun}D. Saunders et al., Nat. Phys. \textbf{6}, 845 (2010).

\bibitem{einstein}A. Einstein, B. Podolsky, and N. Rosen, Phys. Rev.
\textbf{47,} 777 (1935).

\bibitem{rmp}M. D. Reid et al., Rev. Mod. Phys. \textbf{81}, 1727
(2009).

\bibitem{cavalmulti-1}E. G. Cavalcanti et al., Phys. Rev. A \textbf{84},
032115 (2011).

\bibitem{mulitqudits}Q. Y. He et al., Phys. Rev. A \textbf{83}, 032120
(2011).

\bibitem{cvsteerrecent}\textcolor{black}{T. Eberle }\textit{\textcolor{black}{\emph{et
al}}}\textcolor{black}{\emph{.,}}\textcolor{black}{{} }\textit{\textcolor{black}{\emph{Phys.
Rev.}}}\textit{\textcolor{black}{{} }}\textit{\textcolor{black}{\emph{A}}}\textcolor{black}{{}
}\textbf{\textcolor{black}{83}}\textcolor{black}{, 052329 (2011).}\textcolor{red}{}
N. Takei \textit{et al.}, Phys. Rev. A \textbf{74}, 060101(R) (2006)\textcolor{black}{;
A. Samblowski et al., arXiv:1011.5766v2 {[}quant-ph{]}; S. Steinlechner,
arXiv:1112.0461v2 {[}quant-ph{]};} \textcolor{red}{ }\textcolor{blue}{}

\bibitem{onewaysteer}V. Handchen et al., Nature Photonics \textbf{6},
596 (2012).

\bibitem{smithsteerxp}D. H. Smith et al., Nat. Commun.\textbf{ 3},
625 (2012).

\bibitem{bwzeil}\textcolor{black}{B. Wittmann et al., New J. Phys.
}\textbf{\textcolor{black}{14}}\textcolor{black}{, 053030 (2012).}

\bibitem{loopholesteer}A. J. Bennet et al., \textit{\emph{Phys. Rev.
X}} \textbf{2}, 031003 (2012). 

\bibitem{cavalsteer}E. G. Cavalcanti et al., Phys. Rev. A\textbf{.
80}, 032112 (2009).

\bibitem{quantum internet}H. J. Kimble, Nature \textbf{453}, 1023
(2008). 

\bibitem{secretsh-1}A Shamir, Commun. ACM \textbf{22}, 612 (1979).

\bibitem{secretsh} M. Hillery et al., Phys. Rev. A \textbf{59}, 1829
(1999). \textcolor{black}{S. Gaertner et al., Phys. Rev. Lett. }\textbf{\textcolor{black}{98}}\textcolor{black}{,
020503 (2007). A. M. Lance et al., New J. Phys. }\textbf{\textcolor{black}{5}}\textcolor{black}{,
4 (2003).   Y.-A. Chen et al., Phys. Rev. Lett. }\textbf{\textcolor{black}{95}}\textcolor{black}{,
200502 (2005).}

\bibitem{hoftoth}H. F. Hofmann and S. Takeuchi, Phys. Rev. A \textbf{68},
032103 (2003). G. Toth, Phys. Rev. A \textbf{\small 69,}{\small{} 052327
(2004).}{\small \par}

\bibitem{eprr-2}{\small M. D. Reid, Phys. Rev. A }\textbf{\small 40,}{\small{}
913 (1989). }

\bibitem{proofepr} Details are given in supplemental materials.

\bibitem{mermin}N. D. Mermin, Phys. Rev. Lett. \textbf{65}, 1838
(1990). 

\bibitem{olsen tripr}M. K. Olsen et al., Journ Phys B: At. Mol. and
Opt. \textbf{39}, 2515 (2006).

\bibitem{bow} W. Bowen et al., Phys. Rev. Lett. \textbf{90}, 043601
(2003).\end{thebibliography}
\end{document}